Empirical analysis of recent temporal dynamics of research fields:

Annual publications in chemistry and related areas as an example

Lutz Bornmann*$ and Robin Haunschild$

* Science Policy and Strategy Department

Administrative Headquarters of the Max Planck Society

Hofgartenstr. 8,

80539 Munich, Germany.

Email: bornmann@gv.mpg.de

$ Max Planck Institute for Solid State Research

Heisenbergstraße 1,

70569 Stuttgart, Germany.

Email: r.haunschild@fkf.mpg.de, l.bornmann@fkf.mpg.de

**Abstract**

Changes in the number of publications in a certain field might reflect the dynamic of scientific progress in this field, since an increase in the number of publications can be interpreted as an increase in the field-specific knowledge. In this paper, we present a methodological approach to analyse the dynamics of science on lower aggregation levels, i.e., the level of research fields. Our trend analysis approach is able to uncover very recent trends, and the methods used to study the trends are simple to understand for the possible recipients of the results. In order to demonstrate the trend analysis approach, we focused in this study on the annual number of publications (and patents) in chemistry (and related areas) between 2014 and 2020 identifying those fields in chemistry with the highest dynamics (largest rates of change in publication counts). The study is based on the mono-disciplinary literature database CAplus. Our results reveal that the number of publications in the CAplus database is increasing since many years. Research regarding optical phenomena and electrochemical technologies was found to be among the emerging topics in recent years.

# 1  Introduction

One of the basic metrics in scientometrics is the number of publications published by a specific unit (a researcher, an institution, or a country) over a certain period. According to Pinski and Narin (1976) this number "is a measure of total activity only; no inferences concerning importance may be drawn" (p. 298). Changes in the number of publications in a certain field might reflect the dynamic of scientific progress in this field, since an increase in the number of publications can be interpreted as an increase in the field-specific knowledge.

The relationship between number of publications and scientific progress can be assumed, since it is standard in modern science that researchers publish their findings in papers appearing in international peer reviewed journals. "For in a long-standing social reality, only when scientists have published their work and made it generally accessible, preferably in the public print of articles, monographs, and books that enter the archives, does it become legitimately established as more or less securely theirs" (Merton, 1988, p. 620). There are also other ways in science available for knowledge exchange (e.g., presentations or personal discussions); these other ways cannot ensure, however, the attribution of knowledge claims to researchers without any doubt. For van Raan (2005), "bibliometric assessment of research performance is based on one central assumption: Scientists who have something important to say do publish their findings vigorously in the open international journal (serial) literature" (p. 2).

Several studies have been published in scientometrics that report empirical results on the growth of science using the annual numbers of publications from multi-disciplinary databases such as Web of Science (WoS, Clarivate Analytics, or Scopus, Elsevier). An overview of these studies (and the most recent empirical results) can be found in Bornmann and Mutz (2015) and Bornmann, Haunschild, and Mutz (2021). In this paper, we used a methodological approach to analyse the recent temporal dynamics of science on a lower aggregation levels, i.e., the level of research fields. We focused in this study on the annual number of publications in chemistry (and related areas) between 2014 and 2020 identifying those fields in chemistry with the highest dynamics (largest rates of change in publication counts). We statistically analyzed publication (and patent) counts from very recent years for investigating time- and field-specific trends in recent scientific progress. The statistical analyses are based on the philosophy that the number of publications within a field is a good proxy measure for knowledge generation.

## 2 Literature overview and deduction of the methodological trend analysis approach used in this study

In recent years, various researchers (from the scientometrics field) have proposed methods for the investigation of dynamics of research fields. One of the most prominent examples in bibliometrics is the annual reports by Clarivate Analytics (and by the previous Thomson Reuters) on research fronts. The latest analysis from 2020 – conducted together with the Chinese Academy of Sciences – identified emerging specialty areas in scientific research from 2014 to 2019 (Clarivate Analytics and Chinese Academy of Sciences, 2021). The identification of research fronts is mainly based on citations: a "Research Front consists of a core of highly cited papers along with the citing papers that have frequently co-cited the core" (Clarivate Analytics and Chinese Academy of Sciences, 2019, p. 4). The citation analysis in the reports by Clarivate Analytics is based on very short citation windows for identifying currents trends and topics (Moed, 2017). Research front data from Clarivate Analytics has been used by Thomson and Kanesarajah (2017) to determine whether the European Research Council (ERC) has followed its mission to support frontier research.

The overview by Wang (2018) of other approaches for identifying research fronts (besides that of Clarivate Analytics) shows that these approaches are mainly based on citations: citation relations, bibliographic coupling relations, and co-citation relations (see e.g., Boyack & Klavans, 2010; Glänzel & Thijs, 2012; Small, Boyack, & Klavans, 2014). The most recent study by Klavans, Boyack, and Murdick (2020) predicting exceptional growth in research is no exception. Also, methods based on key words have been proposed (see, e.g., Ohniwa, Hibino, & Takeyasu, 2010). The overview by Wang (2018) reveals that only a few studies are based on the number of publications. For example, Bengisu (2003) included these numbers besides publication years in regression models to identify emerging technologies. In this study, we basically followed this approach. In the empirical analysis, Wang (2018) also used annual publication counts as criterion to measure scientific progress. However, following Rotolo, Hicks, and Martin (2015), Wang (2018) additionally used three other criteria. One of these criteria is citation impact.

In this study, we refrained from analyzing citations to identify dynamic research fields in chemistry. We identified three problems in using citations for identifying promising research fields: (a) the current research front is not targeted, since the most recent publications could not be included in the analysis. Short citation windows in bibliometrics do not refer to a few months, but to less than about three years (three years is the minimum citation window in bibliometrics, see Glänzel & Schöpflin, 1995). (b) The results by Wang (2013) show that citation analyses based on short citation windows result in impact measurements that are not valid. Thus, many papers that rapidly received many citations are not necessarily the high-impact papers in the long run. One needs a citation window of many years or decades to reliably identify important papers (Winnink, Tijssen, & van Raan, 2018). (c) Papers including innovative research do not necessarily have high citation impact (in the short run) (Jirschitzka, Oeberst, Göllner, & Cress, 2017). This is why new indicators have been proposed recently that abstained from pure citation counts by combining cited references data for focal and citing papers to identify innovative (disruptive) papers (Bornmann, Devarakonda, Tekles, & Chacko, 2020; Wu, Wang, & Evans, 2019).

With direct citation relations, bibliographic coupling relations, and co-citation relations (Boyack & Klavans, 2010), many studies on emerging research fields and dynamics in research areas have applied complicated statistical techniques and selection criteria. Trained bibliometricians might be familiar with the techniques, but researchers outside of the field – in many cases the recipients of the bibliometric results on research fronts – are mostly not. The loss in transparency and understanding by the use of these techniques (the analyses are more or less black boxes for the recipients) probably leads to a lack of confidence in the results. There is a high risk that researchers who are interested in the identification of fields with high dynamics in their specialist area reject the results, especially in cases in which the results are not completely in agreement with their own subjective judgement. In order to enhance the acceptance of our empirical results in the target community (chemistry and related areas), we tried to use as simple as possible methods; i.e., methods that are well understandable, but lead to meaningful results.

A crucial question in identifying research fields with high dynamics is the choice of the categorization scheme for field- or topic-specific clustering of publications. Previous studies used citation relations, bibliographic coupling relations, and co-citation relations to identify publication clusters (see e.g., Boyack & Klavans, 2010; Glänzel & Thijs, 2012; Small et al., 2014). It is an important disadvantage of these methods that the clusters are not constant: the inclusion of new publications in a database might lead to other clusters of publications, since the pattern of citations changes. Furthermore, the resulting clusters can be numbered but it is very difficult to provide useful labels for the clusters. In a recent attempt, Sjögårde, Ahlgren, and Waltman (2020) tried to solve this problem by comparing various methods for labeling.

Another approach for clustering publications field-specifically is the use of established field-categorization schemes in bibliometrics. Multi-disciplinary classification schemes are available in literature databases such as Web of Science (Clarivate Analytics) or Scopus (Elsevier) and are based on journal sets. The most important advantage of these schemes is that they encompass most of the disciplines in science. However, these schemes also have a number of disadvantages: for Strotmann and Zhao (2010), the subject classifications in these databases are generally not detailed enough … to help with delimiting research fields" (p. 195). Furthermore, many journal sets include multi-disciplinary journals that have a too broad scope: "journals like 'The Lancet' are general medicine journals and as such matching a scheme where 'medicine' is the highest resolution reached. But when a higher resolution is needed, such journals are problematic" (Haddow & Noyons, 2013, p. 1212). To overcome the weaknesses of the multi-disciplinary classification schemes based on journal sets, Waltman and van Eck (2012) proposed an algorithmically constructed classification system based on citation relations covering all disciplines (see also Ruiz-Castillo & Waltman, 2015). The disadvantages of this scheme are, however, that (a) the identified fields are difficult to label, (b) publications without linked citations and linked references are not considered in the scheme, (c) the results are not constant (since new citations might change the results), and (d) many very small clusters (sometimes as small as a single publication per cluster) have to be merged or neglected (Haunschild, Schier, Marx, & Bornmann, 2018).

Although limited to only one discipline and related areas, a promising alternative to the use of multi-disciplinary classification schemes are mono-disciplinary classification schemes based on field-specific databases (Glänzel, Thijs, Schubert, & Debackere, 2009; Wang & Waltman, 2016; Wouters et al., 2015). These classification schemes are available, e. g., for medicine (Medical Subject Headings, see, e.g., Bornmann, Mutz, Neuhaus, & Daniel, 2008) and economics (EconLit, see, e.g., Bornmann & Wohlrabe, 2019). The results by Duffy, Martin, Bryan, and Raque-Bogdan (2008) for PsycINFO – a broad subject-specific database in psychology – show that the coverage of the subject-specific literature is higher in this database than in the multi-disciplinary database Web of Science. In the current study, we focus on chemistry and related areas by the use of the Chemical Abstracts Plus (CAplus$^{SM}$) database produced by the Chemical Abstracts Service (CAS), a division of the American Chemical Society (ACS). The CAplus database is focused on chemistry but also covers related research fields (i.e., parts of biology, medicine, physics, and material sciences). This database has been used in bibliometrics studies, e.g., to field-normalize citations for research evaluation purposes (Bornmann & Daniel, 2008; Bornmann, Schier, Marx, & Daniel, 2011).

The advantages of using the field-categorization scheme in the CAplus database for citation analyses is explained by Bornmann, Marx, and Barth (2013) in detail. The database is the most comprehensive database of publicly disclosed research in chemistry and related areas, and covers publications since 1907 (including the references cited therein since the publication year 1996). The database includes a two-level classification scheme to categorize the publications into five broad headings of chemical research (section headings). These headings are divided in 80 subject areas called Chemical Abstracts (CA) sections (see https://www.cas.org/support/documentation/references/ca-sections). Each publication in the database is assigned to only one CA section according to the main subject field and interest. The assignment of publications to CA sections is intellectually done by indexers on the paper-basis rather than journal-basis.

The results by Braam and Bruil (1992) reveal that the classification of the publications is not affected by so called 'indexer effects': 80% of the publications have been categorized in accordance with the authors' preferences of the corresponding publications. Another advantage of the categorizations in the CAplus database is that papers are usually assigned to only one field, even if these papers are published in multidisciplinary and wide-scope journals. Only some papers are assigned to a cross-section which can be considered as a secondary categorization. In contrast, in the Web of Science, e.g., most journals are assigned to more than one subject category (up to six subject categories). Besides the classification of publications into CA sections, publications are also indexed regarding controlled terms (CT). These CT are very helpful regarding (i) topic searches and (ii) more detailed classification within CA sections. Human indexers (experts in the disciplines chemistry, physics, or biology) classify the publications by using a detailed and controlled vocabulary into CA sections and CTs.

## 3 Methods

This paper introduces methods for studying current dynamics of annual publication counts in research fields. Since this paper is methods-oriented, we abstained from explaining the statistical methods in the methods section, but explain them alongside the presentation of the empirical results on chemistry in section 4. We used chemistry as example in this paper because of mainly two reasons: (i) we have access to the CAplus database with its excellent field-classification scheme (see section 2), and (ii) one of the co-authors (RH) is a chemist by training. However, our methods can be easily transferred to other disciplines with similar high quality schemes. Methodological approaches such as those used in this study have been characterized by Cozzens et al. (2010) as promising approaches to identify emerging technologies: "the first option is to measure growth and characterise growing areas by using existing categories in indexing schemes or terms in controlled indexing vocabularies" (p. 369).

We accessed the CAplus database via the Scientific and Technical Information Network (STN®) International. The content of the database is also available via the application SciFinder®. However, the STN platform provides more detailed search and analysis possibilities. The CAplus database was searched separately for each publication year between 2014 and 2020 for all document types – including patents. Including patents in analyses on output dynamics is important because many inventions are published as patents first and as scientific publications later (Saheb & Saheb, 2020). In the case of patents, only basic patents were counted. This avoids multiple counting of equivalent patents in different jurisdictions. However, if a national patent is very different from other patents in its family, multiple basic patents might exist and might be counted. The resulting publications of all document types were analyzed regarding their CA sections. In the case of the CA section with the highest dynamic (see next section), the publications of this CA section were analyzed regarding their controlled index terms (CT).

The dataset of this study represents publication counts over seven years for 80 CA sections. These CA sections are clustered in the following broad sections: 'Biochemistry' (BIO), 'Organic' (ORG), 'Macromolecular' (MAC), 'Applied' (APP), and 'Physical, Inorganic, and Analytical' (PIA). The results that are presented in the following section are mainly based on statistics recommended by Baldwin (2019). Bengisu (2003) used similar statistics (i.e., regression models) to identify emerging technologies.

# 4 Results

We are interested in this study in identifying CA sections with the largest rate of change across seven recent years (from 2014 to 2020): which CA sections are characterized by large growth rates especially in the most recent years?

## 4.1 Development of publication counts in chemistry

In order to properly frame the findings to the various CA sections in the following, we begin with plotting the development of publication counts across the whole field of chemistry (see Figure 1). The results show an increasing trend between 2000 and 2020. This trend corresponds to the exponential growth for the total scientific literature using data from various databases (Bornmann et al., 2021; Hu, Leydesdorff, & Rousseau, 2020). The publication trend of the WoS research area "Chemistry" is also increasing like the total scientific literature.

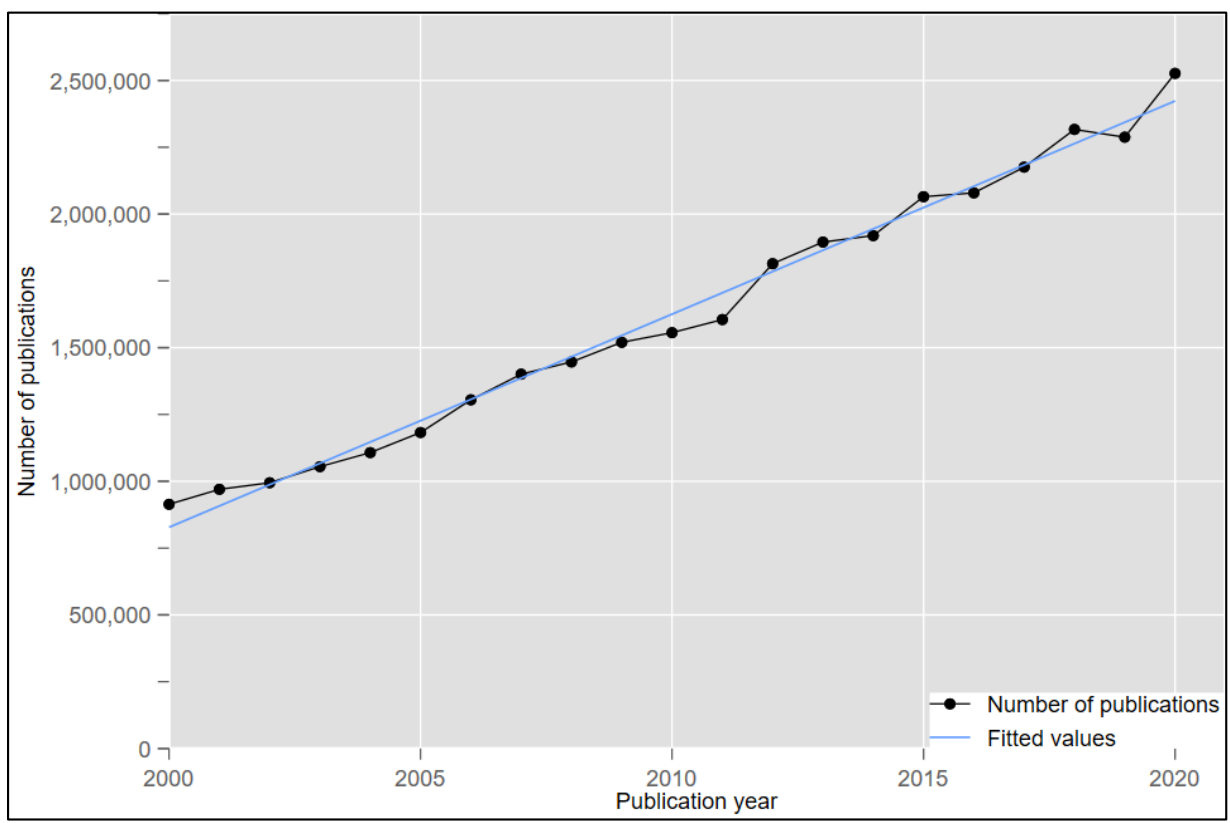

Figure 1. Annual changes in the number of publications in the CAplus database

### 4.2 Development of publication counts in CA sections

Figure 2 shows the number of papers in the various CA sections stratified by publication year. Each dot in the figure represents a CA section outside of adjacent values that are the most extreme values within 1.5 interquartile range of the nearest quartile (Tukey, 1977). The results reveal that several outliers exist in the data, i.e., CA sections with many more papers than the other CA sections. The two CA sections with the largest paper output across the years are 'Pharmacology" and 'Mammalian Pathological Biochemistry', i.e., CA sections in the border areas of chemistry and biology.

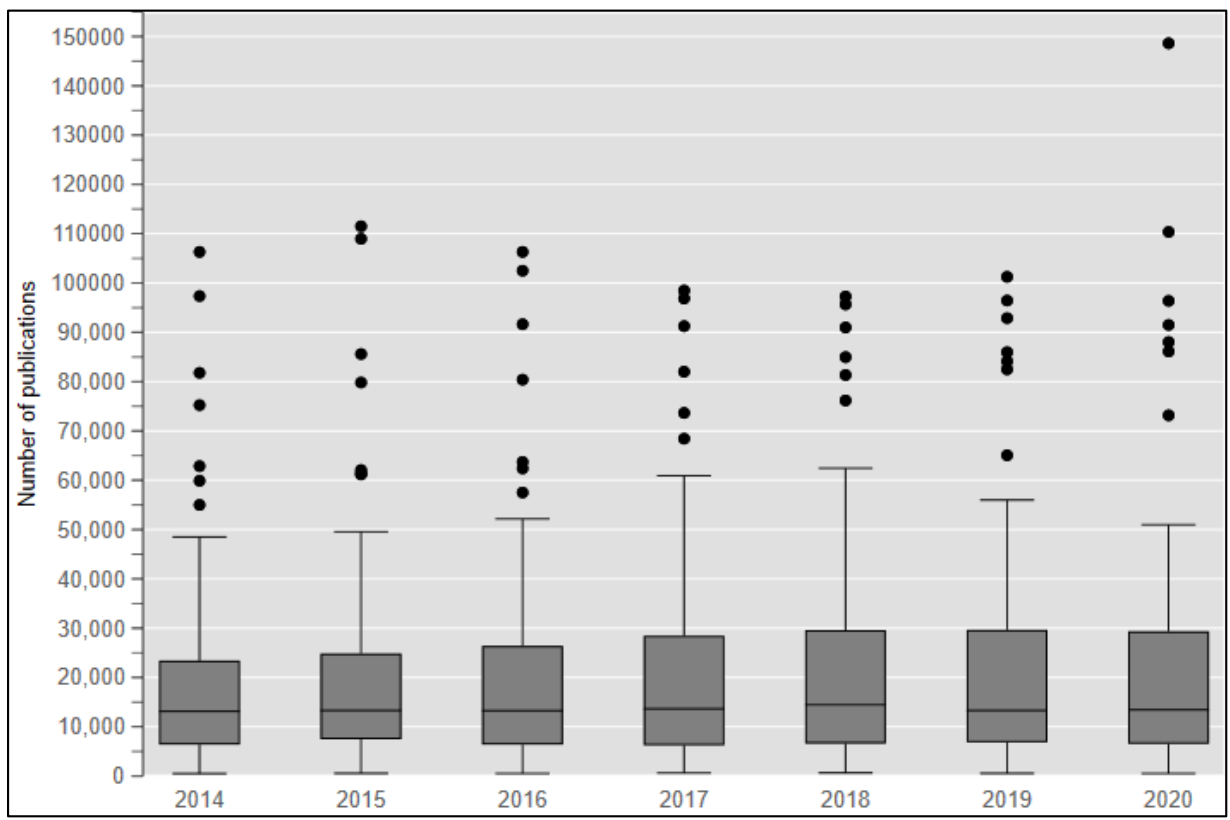

Figure 2. Box plots for the number of papers in 80 CA sections stratified by publication year

Table 1 reports some key figures for the number of papers. The minimum and maximum counts reveal large differences between the CA sections that are also reflected in the large differences between the means and medians. The median counts indicate that the number of publications is relatively constant across all CA sections (however, with an increase in the number of publications in 2017).

Table 1. Key figures for the number of publications in 80 CA sections stratified by publication year

| Publication year | Mean | Median | Standard deviation | Minimum | Maximum |
|---|---|---|---|---|---|
| 2014 | 19,469.61 | 13,130.00 | 21,277.61 | 506 | 106,329 |
| 2015 | 20,487.51 | 13,312.50 | 22,686.52 | 563 | 111,512 |
| 2016 | 20,476.88 | 13,265.50 | 22,483.11 | 530 | 106,326 |
| 2017 | 21,239.95 | 13,662.00 | 22,647.89 | 627 | 98,490 |
| 2018 | 21,942.00 | 14,468.50 | 23,360.36 | 632 | 97,275 |
| 2019 | 21,947.65 | 13,320.50 | 24,011.16 | 565 | 101,263 |

| 2020 | 23,137.39 | 13,433.00 | 27,790.74 | 541 | 148,640 |

In order to analyze how the change of publication counts in CA sections occurs across publication years, we calculated growth-curve models. In Figure 2 and Table 1, we present the results only from year to year considering the whole publication set. Since we are especially interested in how the CA sections change from year to year, we examined the within-CA section change by calculating the mean and variability in change over time. To prepare the data for these analyses, we calculated the differences in the number of publications between the publication years for every CA section. For example, Table 2 shows the corresponding numbers for the CA section 'Pharmacology'. As the differences to the previous years for the CA section reveal, there is no clear increasing or decreasing trend visible.

Table 2. Number of publications assigned to the CA section 'Pharmacology' over seven publication years

| Publication year | Number of publications | Difference to previous year |
|---|---|---|
| 2014 | 106,329 | |
| 2015 | 108,973 | 2644 |
| 2016 | 102,513 | -6460 |
| 2017 | 98,490 | -4023 |
| 2018 | 95,686 | -2804 |
| 2019 | 96,452 | 766 |
| 2020 | 110,376 | 13,924 |

The mean, standard deviation, minimum, and maximum of the differences between the five publication years are reported in Table 3 for all CA sections. The distributions of the differences (separated by publication years) are shown in Figure 3.

Table 3. Key figures for the differences between publication years and the respective previous year for all CA sections

| Publication year | Mean | Median | Standard deviation | Minimum | Maximum |
|---|---|---|---|---|---|
| 2015 | 1,017.90 | 492.00 | 2,397.45 | -1,971 | 14,150 |
| 2016 | -10.64 | -101.00 | 1,845.74 | -6,460 | 6,048 |
| 2017 | 763.08 | 204.50 | 2,676.24 | -9,452 | 11,232 |
| 2018 | 702.05 | 123.50 | 1,934.56 | -3,073 | 7,759 |
| 2019 | 5.65 | 10.50 | 1,465.41 | -3,141 | 7,930 |
| 2020 | 1,189.74 | -0.50 | 5,965.10 | -5,081 | 47,377 |

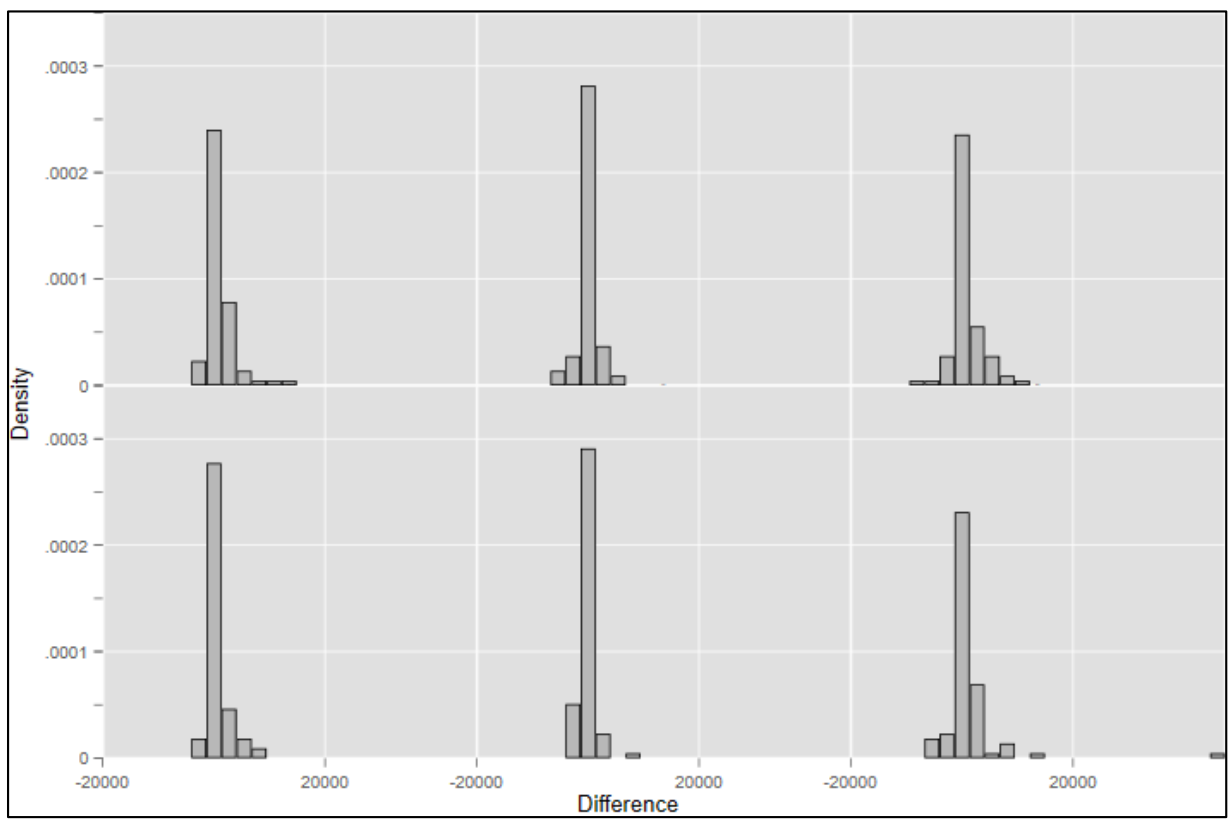

Figure 3. Histograms of annual changes in the number of publications

The distributions in Figure 3 indicate that the differences are rather small in many cases, but the differences are also concerned by a few very large differences. The mean differences in the publication years are negative in one out of six cases (see Table 3): in 2016, the mean difference is -10.64. All key figures in Table 3 reveal that the development over time in the numbers of publications is quite different rather than consistent.

According to Baldwin (2019), descriptive analyses based on the inspection of only two time points are not ideal for studying change. The previous results in this section do not indicate how publication counts changed over the entire range of publication years within CA sections. Change over the complete time range will be addressed therefore in the coming analyses. Figure 4 is a spaghetti plot for the number of publications in the 80 CA sections over time. As the various plots illustrate in the figure, CA sections vary in their pattern of change and vary differently. It seems that hardly any CA section exists with roughly constant numbers. The numbers of publications increase in some CA sections and decrease in many others (which is in agreement with the mean differences reported previously); some CA sections oscillate within the seven years considered in this study.

Figure 4 also includes the total across all CA sections: the graph confirms the previous observation that some CA sections include significantly more publications than the other sections. Furthermore, the starting points (intercepts) of the 80 CA sections in publication counts and their types of changes (slopes) are different.

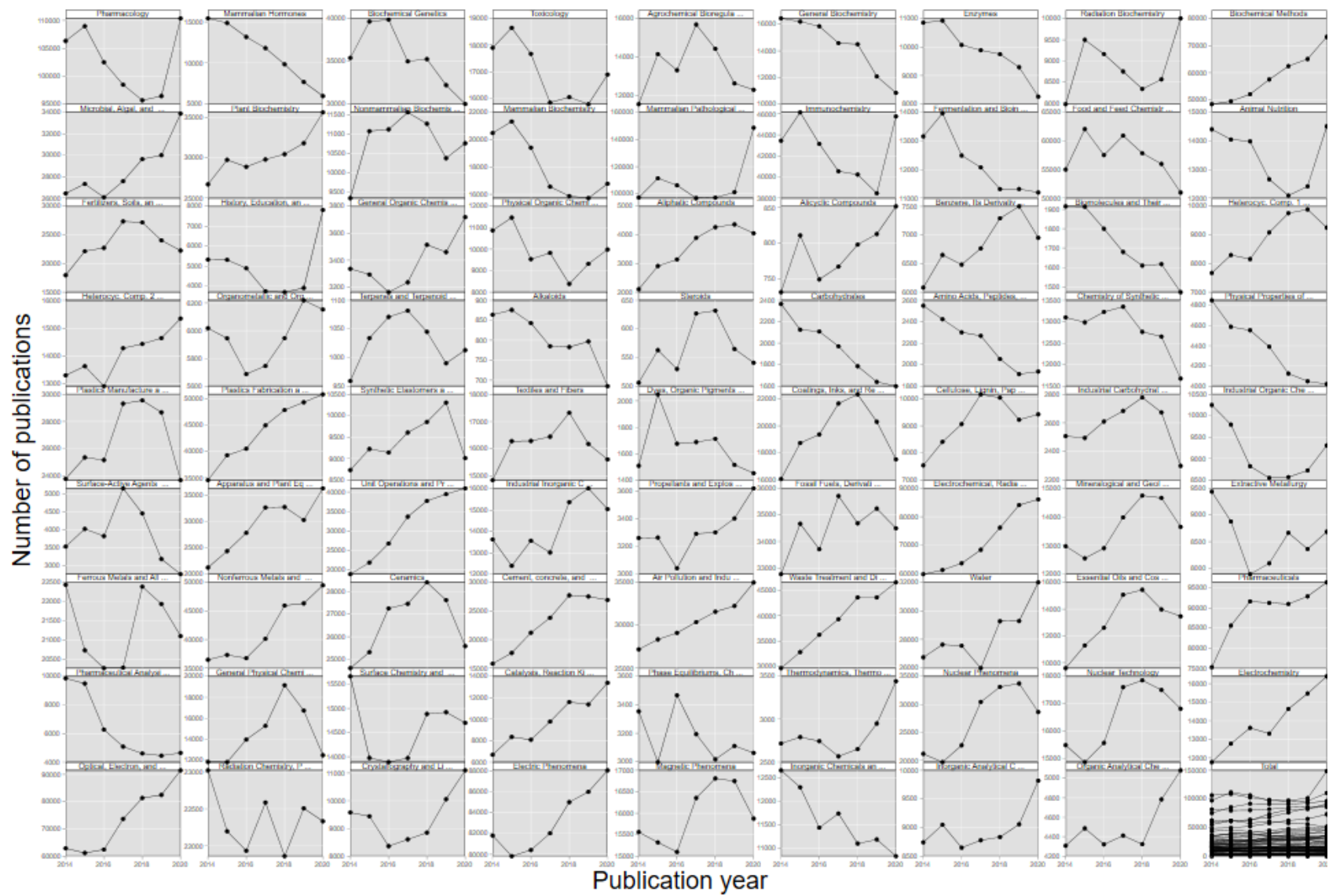

Figure 4. Spaghetti plot showing annual changes in the number of publications for all CA sections

We further explored the variability in the intercepts and slopes by estimating changes in the number of publications for each CA section. We performed a regression analysis for the number of publications for each CA section and collected (analyzed) the intercepts and slopes from the regression models. We fitted an ordinary least squares (OLS) regression for each CA section with number of publications as dependent and time values (transformed publication years) as independent variables

$$P_i = b_0 + b_1 T_i + e_i$$

whereby $i$ is the index for each time unit, $P$ is the number of publications, $T$ is the time value (transformed publication years), and $e$ is the error in prediction. The time variable of this study 'publication year' ranges from 2014 to 2020. The intercept in the regression model is the expected number of publications when the time variable equals 0. Since 'publication year' does not contain this value in our dataset, we transformed the variable whereby 2014 received the value 0 and 2020 the value 6, i.e., $T_i$ = 'publication year' - 2014. This transformation makes $b_0$ the expected value of the number of publications at the baseline (the first publication year considered in this study, i.e., 2014).

We estimated this regression model for each CA section, and saved the intercepts and slopes from each regression to analyze the variability in their relationships.

Table 4. Key figures for the intercepts and slopes from the regression models

| Variable | N | Mean | Standard deviation | Minimum | Maximum |
| --- | --- | --- | --- | --- | --- |
| Intercept | 80 | 19,594.21 | 21,103.69 | 543.68 | 104,802.10 |
| Slope | 80 | 549.60 | 1,347.39 | -1,644.96 | 5,267.32 |

Table 4 shows the key figures for the intercepts and slopes from the 80 regression models. The expected number of publications at the baseline (publication year 2014) – the

intercept – ranges from 543.68 to 104,802.10. The rate of change for the increase in publication years – the slope – ranges from -1,644.96 to 5,267.32; the mean rate of change is positive with around 550. The total expected change across the entire range of publication years (from 2014 to 2020) is 549.60 × 7 = 3,847.20.

Intercept and slope values are (positively) correlated with $r = 0.47$. As the fitted line in Figure 5 shows, the CA sections with high numbers of publications in 2014 are also concerned by high rates of changes with decreasing publication counts.

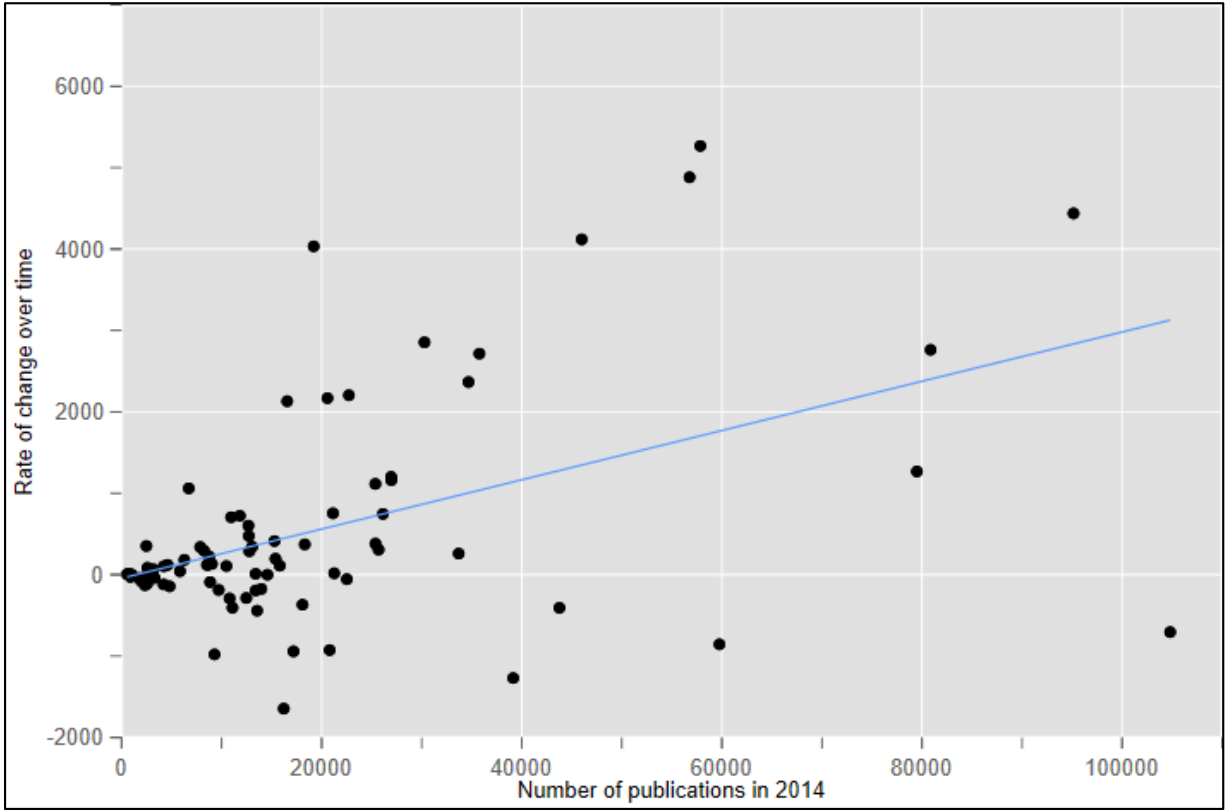

Figure 5. Scatterplot of the CA section-specific intercepts and slopes

We followed the recommendation by Acock (2018) to compute robust standard errors in all regression models. Robust standard errors should be preferred, if not all requirements of the OLS regression model can be fulfilled (e.g., the distributions of the variables do not follow the normal distribution). For robust standard errors, the variance-covariance matrix of the standard errors is estimated using the sandwich estimator. Although the intercepts and slopes did not change by these computations, we needed the adjusted standard errors to calculate confidence intervals for the slopes.

Figure 6 shows the variability in the rate of changes across CA sections with 95% confidence intervals (that are based on robust standard errors and degrees of freedom from the regression models). The rates of change are ranked from low to high on slope. The results show that the CA sections within the broad section 'Biochemistry' mainly show a negative rate of change. The CA sections 'Biochemical Methods', 'Mammalian Pathological Biochemistry', 'Microbial, Algal, and Fungal Biochemistry', and 'Plant Biochemistry' show a positive point estimators but with a 95% confidence interval for 'Mammalian Pathological Biochemistry' reaching into the region of a negative rate of change. The CA sections within the broad sections 'Macromolecular' and 'Organic' mainly show a negligible rate of change, i.e., a constant number of publications. Most CA sections with point estimators with a positive rate of change belong to the broad sections 'Applied' and 'Physical, Inorganic, Analytical'.

'Optical, Electron, and Mass Spectroscopy and Other Related Properties and 'Electrochemical, Radiational, and Thermal Energy Technology' show the highest point estimates of the rate of change. We (exemplarily) used these CA sections for a discussion in the following in more detail. We were interested in hearing more about the topics in the CA sections.

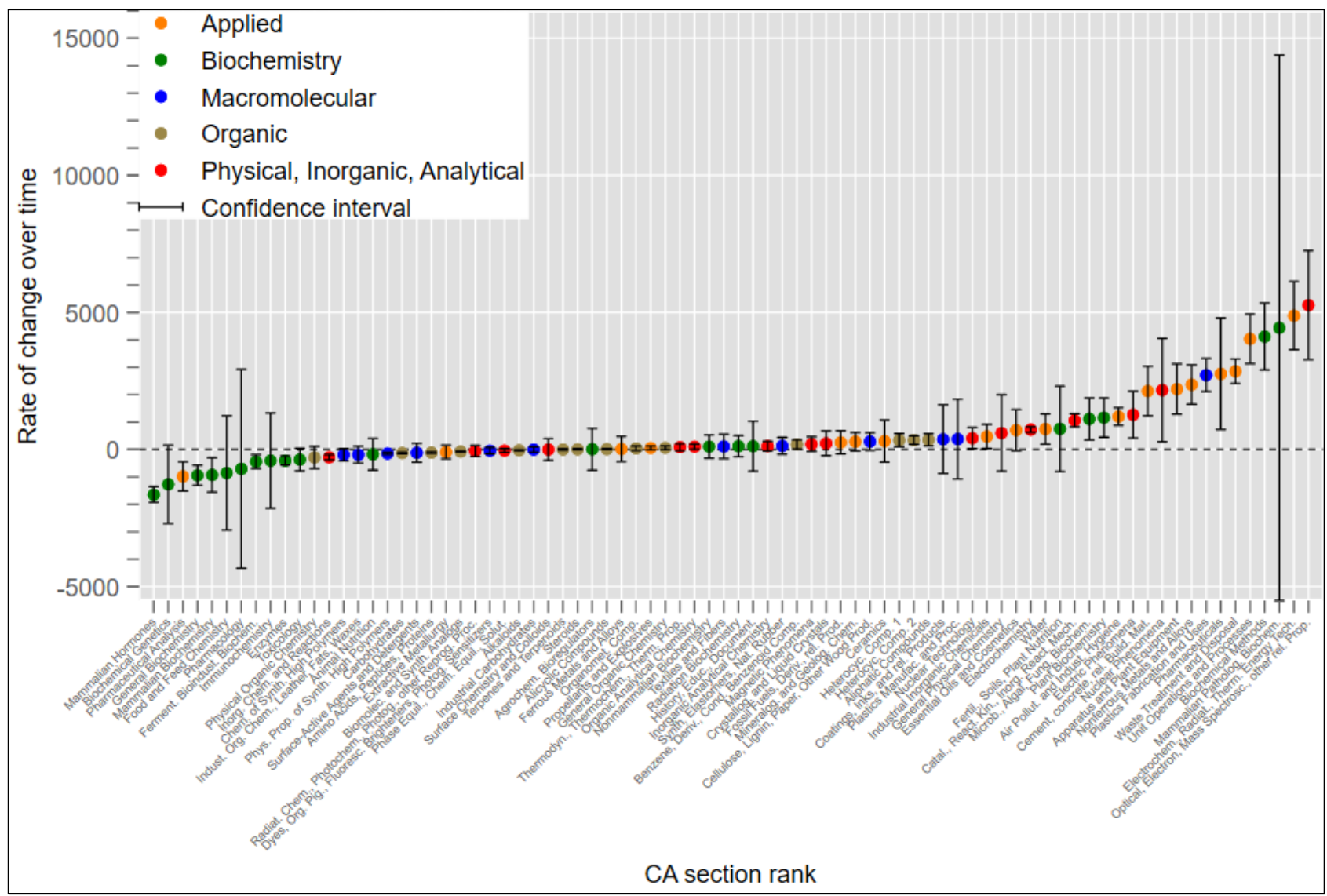

Figure 6. Variability in the rate of change across 80 CA sections (the colors reflect CA main sections)

## 4.3 Publication counts in 'Optical, Electron, and Mass Spectroscopy and Other Related Properties'

We analyzed the publications in 'Optical, Electron, and Mass Spectroscopy and Other Related Properties' regarding their CTs (see section 4.2). The most-current STN tool (STNext) has a limitation of a maximum of 50,000 publications per analysis unit. Between 2014 and 2020, 515,699 publications were assigned to this CA section and 91,580 publications in 2020 alone. Since we are especially interested in the most recent topics, we analyzed the publications assigned to this CA section in 2020 regarding to their CTs. The ten most frequent CTs are shown in Table 7. Note that publications can bear multiple CTs. In total, there are 9,380 unique CTs assigned to publications that are indexed in this CA section.

Table 5. The ten most frequent controlled terms (CTs) of the CA section 'Optical, Electron, and Mass Spectroscopy and Other Related Properties' for publications from 2020

| Controlled term (CT) | Number of publications | In percent |
| --- | --- | --- |
| Photoluminescence | 5,439 | 5.94 |
| Simulation and Modeling | 4,888 | 5.34 |
| UV and Visible Spectra | 4,265 | 4.66 |
| Refractive Index | 4,071 | 4.45 |
| Electroluminescent Devices | 3,981 | 4.35 |
| Lasers | 3,645 | 3.98 |
| Band Gap | 3,506 | 3.83 |
| Surface Structure | 3,377 | 3.69 |
| Polarization | 3,241 | 3.54 |
| Films | 3,210 | 3.51 |
| Total | 91,580 | 100.00 |

About a sixteenth of the publications in the CA section from 2020 bear the CT 'Photoluminescence'. This CT is related to the optical part of the CA section. A slightly lower proportion of the publications bears the CT 'Simulation and Modeling'. As method-related CT, it could be related to all parts of the CA section. Most of the other CTs in Table 7 belong to research on optical phenomena. The large focus on optical technologies in this CA section might be explained with the high relevance of optical phenomena in many fields of research and many industrial applications.

We also analyzed the ten most frequent CTs in Table 7 within this CA section over time. The results are shown in Figure 9 and Figure 10. Figure 9 is a spaghetti plot – similar to the plot presented in Figure 4 – based on CA sections. Figure 10 shows – similar to Figure 6 – the results of 10 regression models using the data on the CTs level (publication counts and time values) as input. The results of the regression analyses reveal that the CT 'Photoluminescence' shows the most dynamic trend over time.

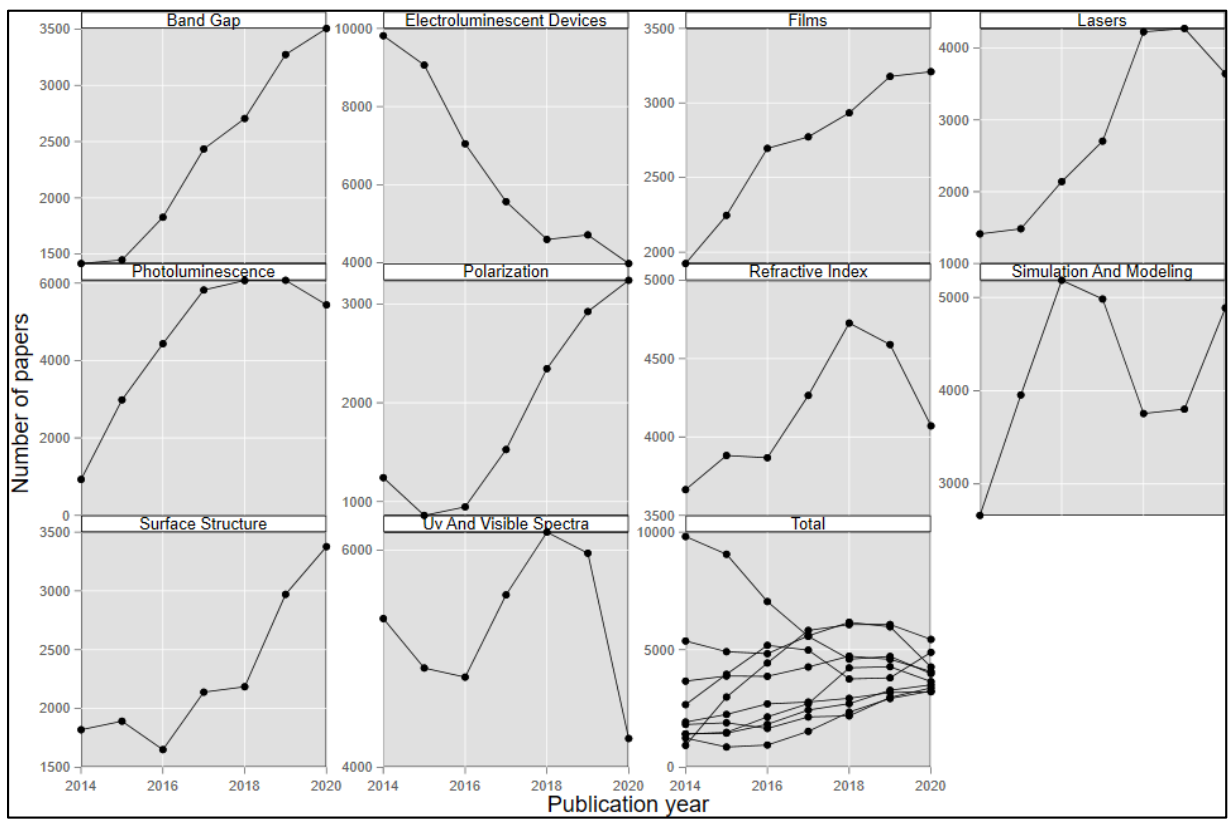

Figure 7. Spaghetti plot showing annual changes in the number of publications for the ten most frequent controlled terms (CTs) within the CA section 'Optical, Electron, and Mass Spectroscopy and Other Related Properties'

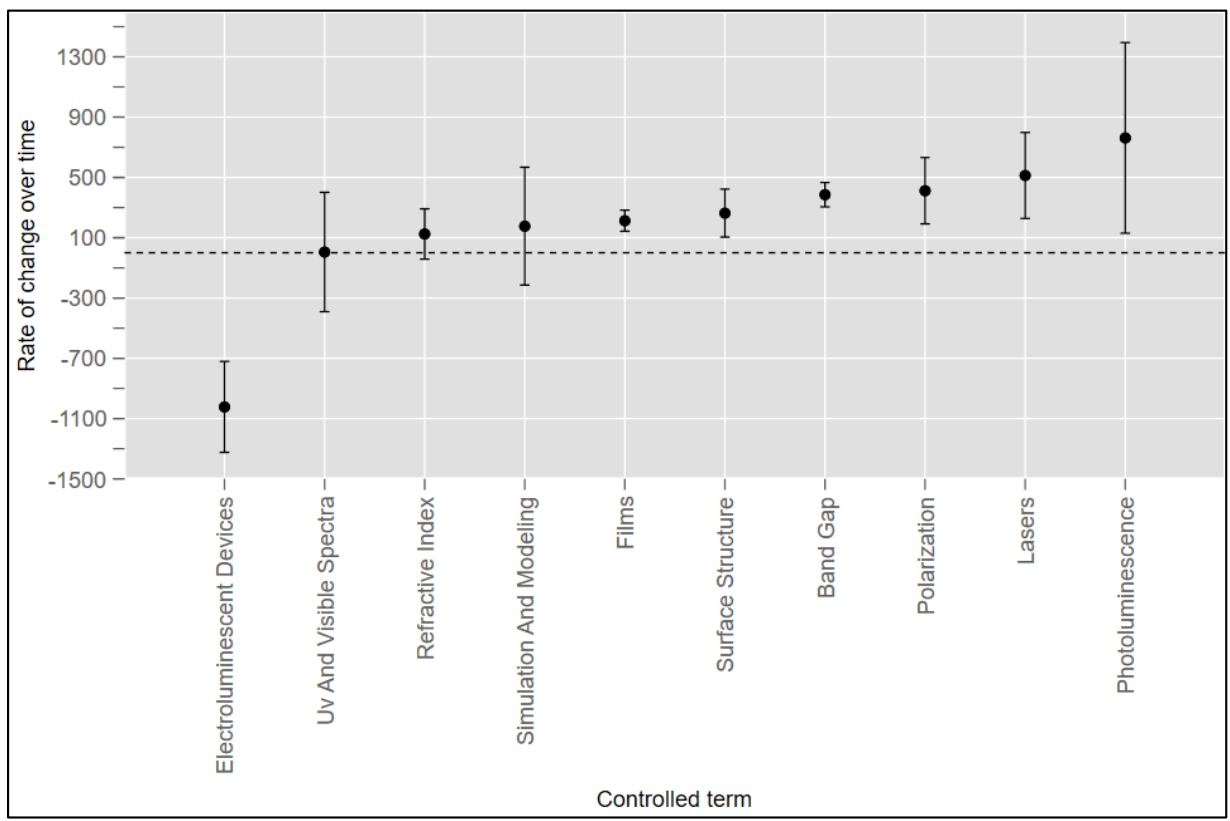

Figure 8. Variability in the rate of change across the ten most frequent controlled terms (CTs) within the CA section 'Optical, Electron, and Mass Spectroscopy and Other Related Properties'

Titles of ten randomly selected publications published in 2020 and indexed under the CA section 'Optical, Electron, and Mass Spectroscopy and Other Related Properties' and the CT 'Photoluminescence' are shown in Table 8. The publications in Table 8 deal with the optical properties of selected substances or materials. Most of the titles include the term 'photoluminescence' or related terms such as 'color centers' (no. 4).

Table 6. Titles of ten randomly selected scientific publications from 2020 and indexed under the CA section 'Optical, Electron, and Mass Spectroscopy and Other Related Properties' and the CT 'Photoluminescence'

| No. | Title |
| --- | --- |

| 1 | Structures and luminescent properties of Eu(III) and Tb(III) complexes with a pentacarboxylic acid |
|---|---|
| 2 | Effect of annealing on photoluminescence properties in zirconium dioxide nanotubes |
| 3 | Bandgap and PL Spectra of Bismuth Ferrite-Nickel Ferrite Nanocomposites Synthesized by Wet Chemistry |
| 4 | High-Temperature Spin Manipulation on Color Centers in Rhombic Silicon Carbide Polytype 21R-SiC |
| 5 | Correlation between the internal quantum efficiency and photoluminescence lifetime of the near-band-edge emission in a ZnO single crystal grown by the hydrothermal method |
| 6 | Semi-organic nonlinear optical material: (((4-sulfonatophenyl)ammonio)oxy)zirconium for dielectric and photonics applications |
| 7 | The Influence of the Parameters of a Short-Period InGaAs/InGaAlAs Superlattice on Photoluminescence Efficiency |
| 8 | Interplay of Multiexciton Relaxation and Carrier Trapping in Photoluminescent CdS Quantum Dots Prepared in Aqueous Medium |
| 9 | Preparation and novel photoluminescence properties of the self-supporting nanoporous InP thin films |
| 10 | Photoluminescence Investigation of Carrier Localization in Colloidal PbS and PbS/MnS Quantum Dots |

## 4.4 Publication counts in 'Electrochemical, Radiational, and Thermal Energy Technology'

We also investigated the publications in 'Electrochemical, Radiational, and Thermal Energy Technology' (see section 4.2). Between 2014 and 2020, 500,016 publications were assigned to this CA section and 86,209 publications in 2020 alone. We focused on the CTs of the publications assigned to 'Electrochemical, Radiational, and Thermal Energy Technology' in 2020. In total, 10,701 distinct CTs were assigned to the publications. Table 7 presents the ten most frequent CTs.

Table 7. The ten most frequent controlled terms (CTs) of the CA section 'Electrochemical, Radiational, and Thermal Energy Technology' for publications from 2020

| Controlled term (CT) | Number of publications | In percent |
|---|---|---|
| Lithium-Ion Secondary Batteries | 13,172 | 15.28 |
| Electric Current-Potential Relationship | 9,876 | 11.46 |
| Surface Structure | 8,380 | 9.72 |
| Battery Anodes | 8,121 | 9.42 |

| Battery Cathodes | 7,934 | 9.20 |
|---|---|---|
| Surface Area | 6,684 | 7.75 |
| Current Density | 6,529 | 7.57 |
| Fluoropolymers | 6,309 | 7.32 |
| Nanoparticles | 6,208 | 7.20 |
| Carbon Black | 5,230 | 6.07 |
| Total | 86,209 | 100.00 |

About a seventh of the publications bear the CT 'Lithium-Ion Secondary Batteries'. About a tenth of the publications have the CTs 'Electric Current-Potential Relationship' and 'Surface Structure'. The CT 'Surface Structure' might be relevant to all parts of this CA section whereas the other two CTs are related to the electrochemical part of this CA section. The most of the other CTs in Table 7, e.g., 'Battery Anodes', 'Battery Cathodes', and 'Current Density' belong to research on electrochemical technologies. This large focus on electrochemical technologies in this CA section might be explained with research on renewable energies and a connected high demand on chemical energy storage.

We also investigated the ten most frequent CTs in Table 7 over time. The results are presented in Figure 9 and Figure 10. The spaghetti plot in Figure 9 is based on CA sections. Figure 10 reveals the results of 10 regression models based on data on the CTs level (publication counts and time values) as input. The results of the regression analyses reveal that the CT 'Lithium-Ion Secondary Batteries' shows the most dynamic trend over time.

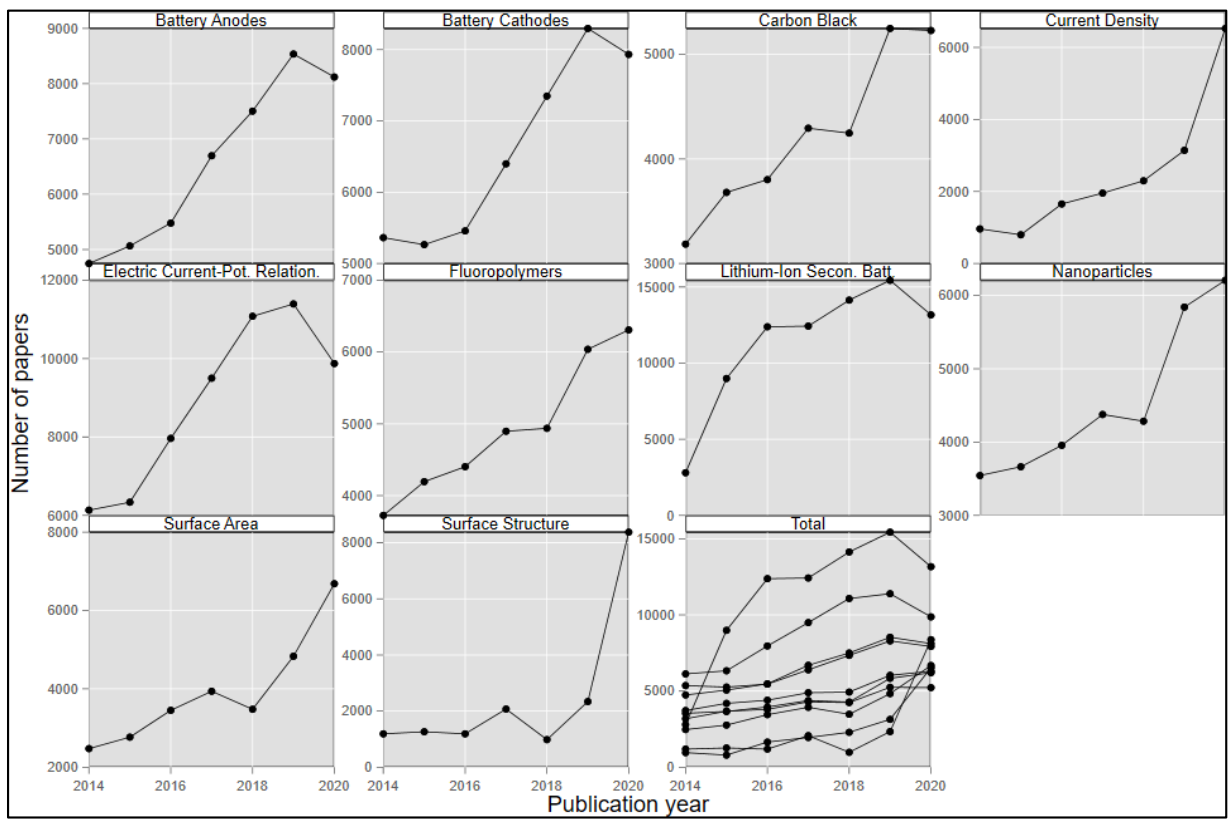

Figure 9. Spaghetti plot showing annual changes in the number of publications for the ten most frequent controlled terms (CTs) within the CA section 'Electrochemical, Radiational, and Thermal Energy Technology'

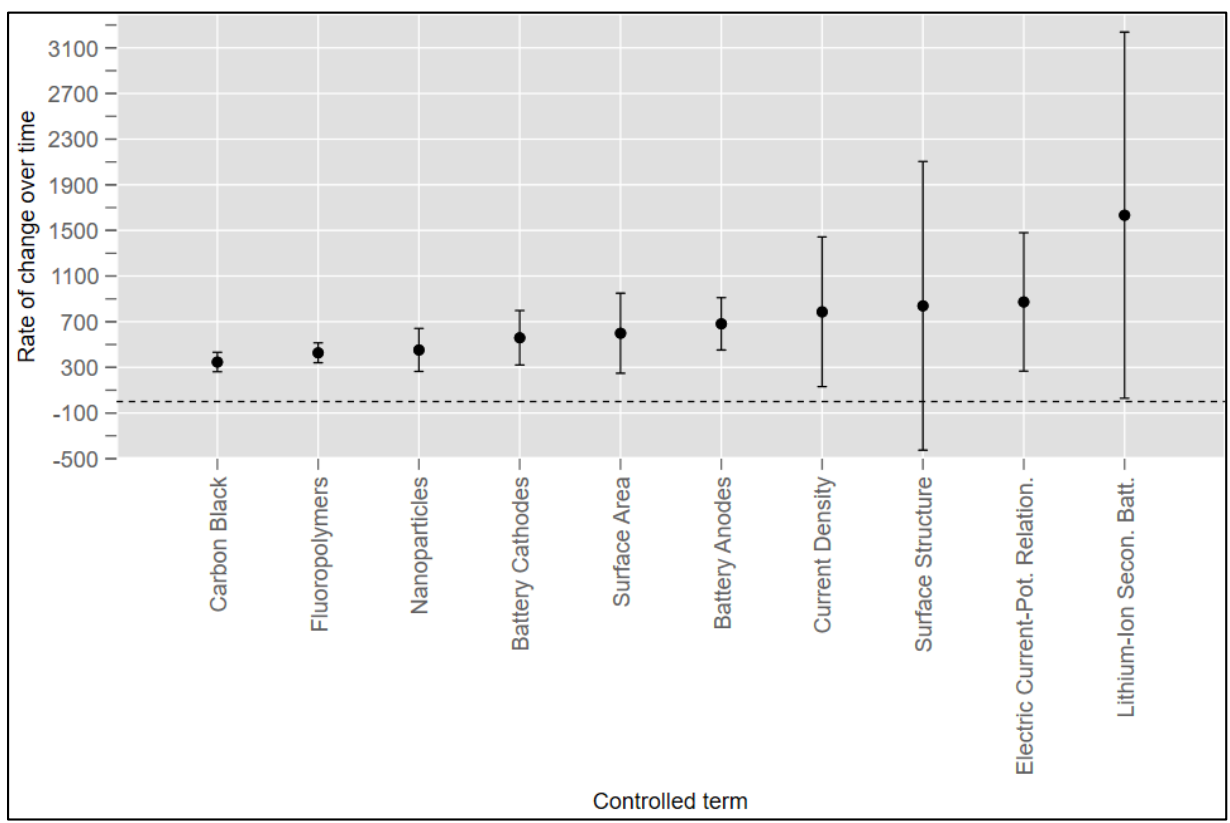

Figure 10. Variability in the rate of change across the ten most frequent controlled terms (CTs) within the CA section 'Electrochemical, Radiational, and Thermal Energy Technology'

Titles of ten randomly selected publications published in 2020 and indexed under the CA section 'Electrochemical, Radiational, and Thermal Energy Technology' and the CT 'Electric Current-Potential Relationship' are shown in Table 8. The publications in Table 8 deal with batteries, mainly their electrode material (numbers 1, 3-7).

Table 8. Titles of ten randomly selected scientific publications from 2020 and indexed under the CA section 'Electrochemical, Radiational, and Thermal Energy Technology' and the CT 'Electric Current-Potential Relationship'

| No. | Title |
|---|---|
| 1 | Combining $Zn_{0.76}Co_{0.24}S$ with S-doped graphene as high-performance anode materials for lithium and sodium-ion batteries |
| 2 | Preparation of new composite electrolytes for solid-state lithium rechargeable batteries by compounding LiTFSI, PVDF-HFP and LLZTO |

| 3 | Key Parameter Optimization for the Continuous Synthesis of Ni-Rich Ni-Co-Al Cathode Materials for Lithium-Ion Batteries |
|---|---|
| 4 | Improving the electrochemical performance of a natural molybdenite/N-doped graphene composite anode for lithium-ion batteries via short-time microwave irradiation |
| 5 | Exploring the Role of Vinylene Carbonate in the Passivation and Capacity Retention of $Cu_2Sb$ Thin Film Anodes |
| 6 | Flocculant Cu Caused by the Jahn-Teller Effect Improved the Performance of Mg-MOF-74 as an Anode Material for Lithium-Ion Batteries |
| 7 | Lithium-silicon compounds as electrode material for lithium-ion batteries |
| 8 | Multi-physics equivalent circuit models for a cooling system of a lithium ion battery pack |
| 9 | Synthesis of the Se-HPCF composite via a liquid-solution route and its stable cycling performance in Li-Se batteries |
| 10 | Overcharge and aging analytics of Li-ion cells |

# 5 Discussion

Bibliometrics is used today in very different contexts. Most of the applications refer to the evaluation of researchers and research institutions with regard to their past performance (output and citation impact). The use of bibliometrics to analyze trends, dynamics, or emergence of research topics has also been undertaken hitherto, but the number of studies is comparably low. The lower number probably lies in the difficulty of the application: one needs a literature database that covers nearly all relevant publications (e.g., in a discipline) and a method that is able to uncover very recent trends. The method which is used to study the trends should be as simple as possible (to be understandable to the recipients of the results). If the method used for the analyses needs data referring to a period many years ago, the results cannot be used to uncover recent dynamics. If methods are too complex and scarcely understandable by the recipients of the trend analyses, the recipients might distrust the results.

In our study, we tried to develop a trend analysis approach that fulfills our requirements. With the CAplus database, we used a mono-disciplinary literature database with an excellent coverage of the literature in chemistry (and related areas). The field-classification system used in the database reliably and validly assigns the literature in chemistry to fields. We abstained from using citations in our trend analysis approach, since impact measurements should be based on citation windows including several publication years. With descriptive statistics and linear regression models, we used simple standard statistics that are well understandable. Recipients of the results should be in the position to understand the statistics.

Overall, the number of publications in the CAplus database is increasing over recent years. Mainly, research regarding photoluminescence and lithium-ion secondary batteries have been increasing with high dynamics between 2014 and 2020. Photoluminescence is the process in which photons are absorbed by materials or compounds that afterwards emit photons usually with a different wave length. The system switches into an electronically excited state upon absorption of the incoming photon and relaxes into the ground state upon emission of the outgoing photon. Other relaxation processes often occur in-between. Therefore, the outgoing photon has usually a larger wave length (i.e., less energy) than the incoming photon. Such processes are important in many optical phenomena. One very important application of photoluminescence is photoluminescence spectroscopy. In contrast to many other spectroscopic methods, photoluminescence spectroscopy is non-destructive because only photons are used for probing. This makes photoluminescence spectroscopy a very versatile method that is generally applicable to systems that exhibit luminescence.

Efficient batteries are needed for storing energies from renewable sources. Although the Shockley Queisser limit (Shockley & Queisser, 1961) provides an upper bound for solar cells, in principle, the sun provides more than enough energy for covering current demands (Khare, 2020). The most severe disadvantages of solar cells are rather low energy conversion efficiency due to optical loss and lack of sufficient storage capacity of produced energy. Numerous studies that tried to provide improvements regarding the optical loss problem have been published. Kim, Lee, and Kwak (2020), Hadadian, Smått, and Correa-Baena (2020), and Khare (2020) provide reviews of recent developments in solar cell research from different perspectives. The problem of insufficient energy storage capabilities is being tackled by battery research.

Thus far, the main batteries used are lithium-ion batteries due to their high output voltage, longevity in the life cycle, high energy density, low self-discharge rate, and wide operating temperature range (Tian, Qin, Li, & Zhao, 2020). However, the performance of lithium-ion batteries severely degrades due to aging and improper operation. Degradation of battery performance can lead to battery leakage, partial short circuit problems, and insulation damage. Some cell phones with lithium-ion batteries (e.g., iPhone, see https://www.ainonline.com/aviation-news/aviation-international-news/2012-02-01/battery-fires-keeping-li-ion-caged) spontaneously caught fire.

Major improvements of current and development of better electrode materials is essential to meet the increasing energy storage demands. Such major improvements might be possible by gaining a deeper understanding of electrode reaction processes, degradation mechanisms, and thermal decomposition mechanisms under realistic operation conditions (Kim, Choi, et al., 2020; Liu et al., 2019). All-solid-state batteries have emerged as a very promising new generation of energy storage materials for mobile applications (including electric cars) due to their potential to exhibit high safety, high energy density, and long cycle life (Janek & Zeier, 2016; Zhang et al., 2018). Redox flow batteries seem to be very attractive for stationary energy storage applications (Wu, Dai, Zhang, & Li, 2020). Ongoing efforts for diminishing the disadvantages of current energy storage technologies might be one of the reasons for the findings in our study.

What are the limitations of our study? Our trend analysis approach is based on research that appeared in publications (especially papers published in journals). What we lost in the analyses of dynamics are other forms of communicating research (e.g., presentations or blogs). The research paper might be the most important form, but these other forms also exist. Another disadvantage of our study concerns the unweighted use of publications counts. As

van Raan (2005) outlines "journal articles …are not equivalent elements in the scientific process; they differ widely in importance" (p. 2). Possible improvements of our trend analysis approach might focus therefore on methods that consider these differences in publications. We imagine that the importance could be considered by including journal metrics in the approach. We expect that – on average – the more reputable journal publishes the more important papers. However, unweighted publication counts probably are only a slight limitation of our work because differences in importance between research papers do not preclude that they can be roughly equivalent in terms of the amount of knowledge generation.

We recommend using our trend analysis approach with a mono-disciplinary database (to have the best possible coverage of the literature and field-classification system). The disadvantage of using mono-disciplinary instead of multi-disciplinary databases might be the possible neglect of research that takes place between disciplines. This depends, however, on the specificity of the database: whereas the CAplus database covers neighbor disciplines such as materials sciences and physics relatively well, the coverage of chemistry in Inspec (a comprehensive engineering research database; see https://www.elsevier.com/solutions/engineering-village/content/inspec) or WoS is relatively worse.

We used linear regression models for the trend analyses in this study, because we are basically interested in an increasing or decreasing trend over time. The specific distribution function was not the focus of our interest. However, we think the distribution function could be an interesting point for the further development of our trend analysis method. For example, with segmented regression models the temporal distribution can be split up into segments and linear regressions can be performed per segment (see, e.g., Bornmann & Mutz, 2015).

The statistical trend analysis approach is only the first step in the analysis of recent temporal dynamics of research fields. The results should be interpreted against the backdrop of expert knowledge in the second step. This is one of the reasons why a chemist has been involved in the study.

# 6 Conclusions

We introduced a methodological approach to analyse the dynamics of science on the level of research fields. Our approach is able to uncover very recent trends, and the methods applied to investigate the trends are simple to understand. To reveal the trend analysis approach in this study, we analysed annual numbers of publications in chemistry (and related areas) between 2014 and 2020. We identified those CA sections in chemistry with the highest dynamics. These are those CA sections with the largest rates of change in publication counts. This study is based on CAplus. In principle, the trend analysis approach that we propose in this paper can be used in all research fields in which bibliometrics can be applied.

# Acknowledgements

The bibliometric data used in this paper are from the CAplus database via the Scientific and Technical Information Network (STN®) International.